\documentclass[10pt]{article}
\usepackage{graphicx}
\usepackage{dcolumn}
\usepackage{bm}
\usepackage{fixmath}
\usepackage{amssymb,amsthm,amsmath}
\usepackage{xcolor,paralist,hyperref,titlesec,fancyhdr,etoolbox}
\usepackage[margin=2.54cm]{geometry}
\usepackage{authblk}
\usepackage{indentfirst,csquotes}

\DeclareMathOperator{\csch}{csch}

\begin{document}

\title{\textbf{The magnetic inverse problem for two stacked layers of sources}}
\author[1,2]{M. T. M. Woodley\footnote{mw2970@bath.ac.uk}}
\author[2]{T. Coussens}
\author[3]{W. Evans}
\author[2,4]{M. Withers}
\author[2]{L. Page}
\author[2]{D. Nightingale}
\author[2]{D. Nicolau}
\author[4]{G. Kendall}
\author[2]{F. Oru\v{c}evi\'{c}}
\author[2,3]{P. Kr\"{u}ger} 
\affil[1]{\small\textit{Centre for Photonics, Department of Physics, University of Bath, Bath, BA2 7AY, UK.}}
\affil[2]{\small\textit{Department of Physics and Astronomy, University of Sussex, Brighton, BN1 9QH, UK.}}
\affil[3]{\small{\textit{Physikalisch-Technische Bundesanstalt (PTB), Abbestr. 2-12, 10587 Berlin, Germany.}}}
\affil[4]{\small{\textit{CDO2 Ltd., London House, High Street, Mayfield, TN20 6AQ, UK.}}}
\date{}

\maketitle

\begin{abstract}
We present calculations that reconstruct electronic current densities in two stacked layers at known depths, using magnetic field data. Solving this inverse problem requires knowledge of the magnetic field in two planes -- one above both current layers, one below -- corresponding to non-invasive measurements of the field. We corroborate the accuracy of current density reconstruction from the resulting system of equations using a numerical simulation. This method is anticipated to be applicable to non-destructive current imaging for quality assurance in a range of applications featuring two-layer geometries, including printed circuit boards, capacitors, fuel cells, and battery cells; we focus particularly here on battery cells, due to their rapidly increasing relevance for automotive applications. This method also offers a framework for generalising the model to more than two layers in future work.
\end{abstract}

\section{Introduction}
Inverse problems are very general in the mathematical sciences, with applications ranging from medical imaging \cite{Bertero,Carpio2008,Greensite} to geophysics \cite{Blakely1995,Richter2020, Snieder1999}. This article is concerned with solving the magnetic inverse problem, in which electric current density is reconstructed from the magnetic field that it produces. It is often especially useful to consider non-invasive measurements of the magnetic field, whereby data are taken only outside of the current-carrying volume under consideration. This is because, in many cases, invasive measurements would destroy, or at least in some way compromise, that current source.

The underlying physics of an inverse problem may be understood through its corresponding forward problem: formulating and using a model to predict observable data. Here, the relevant model is the Biot-Savart law, in which current density is used to calculate a resulting magnetic field. Inverse problems are often ill-posed, as defined by Hadamard \cite{Hadamard1902}. Indeed, it has previously been reported that the general problem of inverting the Biot-Savart law, in order to reconstruct a three-dimensional current source from magnetic field data, is ill-posed, due to non-uniqueness -- see Ref.~\cite{Lima2006}, for example. However, it has been demonstrated by Roth et al. \cite{Roth1989} that, if one of the components of the current density is neglected, then this inverse problem may be solved uniquely using two-dimensional Fourier transforms and spatial filtering. The procedure outlined by Roth et al. assumes that the thickness of the current layer involved is much smaller than the separation between that layer and the measurement of the field. This assumption has also been made in more recent studies, including investigations of current density mapping in thin films of superconductors \cite{Brandt1995,Grant1994,Pashitski1997}, and of electron transport in graphene \cite{Ku2020}. However, this thin current layer requirement was relaxed in the more general treatments of Jooss et al. \cite{Jooss1998} and Zuber et al \cite{Zuber2018}.

There is considerable interest in using current density reconstruction methods for the characterisation of electronic and electrochemical devices, such as, respectively, printed circuit boards \cite{Roth1989} and battery cells, particularly for the detection of hot-spots or defects prior to failure \cite{Bason2022,Brauchle2023}. Using magnetometers in conjunction with such reconstruction techniques offers the possibility of non-destructive testing and real-time imaging capabilities \cite{Bason2022}. 

The problem of reconstructing current densities in two stacked layers is addressed in this article. Exact expressions for the current densities are provided, assuming that magnetic field data are taken both above and below the two current layers. Our approach also forms a basis for further work involving stacks of more than two current layers. We mainly consider battery cells here, but this approach should also be applicable to a range of other applications, including printed circuit boards, capacitors, and fuel cells. Since certain types of battery cell feature two flat, separated current collectors \cite{Garayt2023,Murray2019,Pathan2019}, it is anticipated that this work would be beneficial for the assessment of such cells, either at the end of a production line or as part of a maintenance programme. This assessment could be carried out using whichever type of magnetometer is most appropriate for the application in question, depending on requirements of sensitivity, dynamic range, length-scale, etc. One of these types is the flux-gate. These devices work by producing a voltage in response to the asymmetrical magnetic saturation of a ferromagnetic core in the presence of an external magnetic field \cite{Janosek2016}. Flux-gates can achieve noise densities of a few $\mathrm{pT}/\sqrt{\mathrm{Hz}}$ \cite{Priftis2024}. By contrast, sensors that operate using quantum-mechanical principles can achieve much greater sensitivity. These include: superconducting quantum interference devices (SQUIDs), which use the discrete nature of magnetic flux and the Josephson effect to modulate the electronic properties of a loop of superconductor \cite{Chesca2004}; and optically pumped magnetometers (OPMs), which use laser light to probe a magnetic resonance of an atomic ensemble (often an alkali metal vapour) in order to infer an external magnetic field. Examples of SQUIDs and OPMs can achieve noise densities of the order of $\mathrm{fT}/\sqrt{\mathrm{Hz}}$ \cite{Buchner2018,Colombo2016,Coussens2024}, and it is these types of sensor that are of particular interest for the non-invasive imaging of electrochemical phenomena, especially in battery cells.

\section{Reconstructing current densities in two stacked layers}

\subsection{The Biot-Savart law}
It is assumed here that all magnetic fields involved, and the current densities that give rise to them, are static in time. This is not a severe restriction, since the quasi-static approximation may be used in many cases \cite{Jackson2009}, provided any dynamics present are sufficiently slow -- i.e., \textit{the system is small compared with the electro-magnetic wavelength associated with the dominant time-scale of the problem} \cite{Larsson2007}. Indeed, the quasi-static approximation is more accurate than expected, since any time dependence in the current density cancels out to first order \cite{Griffiths2017}. Consequently, we assume that the Biot-Savart law holds:

\begin{equation}\label{eq:B-S}
    \mathbold{H(r)} = \frac{1}{4\pi}\iiint\limits_{V}\frac{\mathbold{J(r')}\times(\mathbold{r}-\mathbold{r'})}{|\mathbold{r}-\mathbold{r'}|^{3}}d^{3}\mathbold{r'}.
\end{equation}

\noindent Here, $\mathbold{r}$ and $\mathbold{r'}$ are position vectors: the former gives the location of a measurement of the magnetic field, $\mathbold{H(r)}$, outside the current-carrying region; the latter is a variable with which to integrate contributions from the electric current density, $\mathbold{J(r')}$, over the volume, $V$, containing the current-carrying region. All vectors here are three-dimensional. The general non-uniqueness of calculating $\mathbold{J}$ can be appreciated from the fact that, when produced by bounded currents, the $\mathbold{H}$ in a current-free region can be derived from a scalar potential satisfying Laplace's equation, which admits the possibility of adding or subtracting source distributions that have no effect on the resulting field \cite{Lima2006}. The connection to Laplace's equation and its solutions (harmonic functions), links the magnetostatic inverse problem closely to the study of potential fields \cite{Blakely1995}. An important general property of potential fields is that one can use the field at one location in space to calculate the field at a different location, via so-called upward or downward continuation \cite{Ravat}. This property will be exploited in this article.

\newpage

\subsection{The layout of the problem}
\begin{figure}
    \centering
    \includegraphics[width=0.8\textwidth]{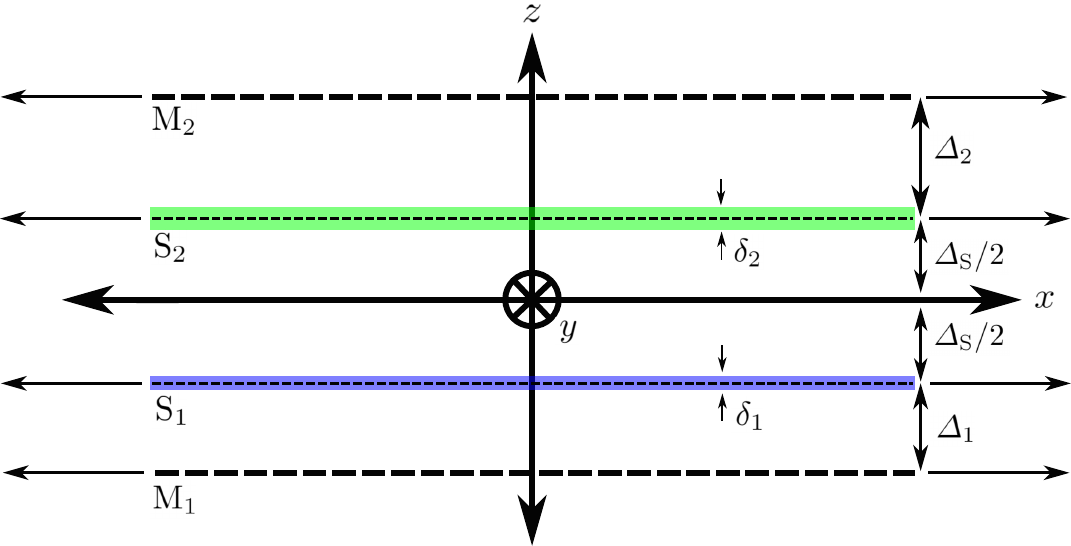}
    \caption{A schematic of the problem, featuring two current layers, $\mathrm{S}_{1,2}$, respectively of thickness $\delta_{1,2}$, and two measurement planes, $\mathrm{M}_{1,2}$, all parallel with the $x$-$y$-plane and stacked in the $z$-direction. The separations between the current layers and measurement planes are measured with respect to the middle (in $z$) of each current layer. In this figure, $\delta_{2}>\delta_{1}$, but this is only for the sake of example and to emphasise that they need not be equal. It is implicit that $\delta_{1,2}<\Delta_{\mathrm{S}}$, but no other constraints are placed on $\delta_{1,2}$ within the model. All layers and planes are considered to be infinitely extended in $x$ and $y$, prior to performing spatial filtering on the problem. However, non-zero current densities will only be localised to small portions of $\mathrm{S}_{1,2}$.}
    \label{fig:layout}
\end{figure}

We will use Cartesian coordinates and consider two layers of current, $\mathrm{S}_{1,2}$ -- generally of different thicknesses, $\delta_{1,2}$, and with current densities $\mathbold{J}_{\mathrm{S}_{1,2}}(\mathbold{r}_{1,2}')$, each respectively generating magnetic fields $\mathbold{H}_{1,2}(\mathbold{r}_{1,2})$ -- and two measurement planes, $\mathrm{M}_{1,2}$, with one plane above the current stack, and one below, as illustrated in Fig.~\ref{fig:layout}. $\mathrm{S}_{1,2}$ are situated at $z=z_{\mathrm{S}_{1,2}}$ and $\mathrm{M}_{1,2}$ at $z=z_{\mathrm{M}_{1,2}}$. The $x$-$y$-plane is sandwiched equidistantly between $\mathrm{S}_{1}$ and $\mathrm{S}_{2}$, and all measurement planes and current layers are parallel with the $x$-$y$-plane and stacked in the $z$-direction. The following positive differences in $z$, between the current layers and the measurement planes, may be defined: $\Delta_{\mathrm{S}} := z_{\mathrm{S}_{2}} - z_{\mathrm{S}_{1}} = 2|z_{\mathrm{S}}|$, $\Delta_{1} := z_{\mathrm{S}_{1}} - z_{\mathrm{M}_{1}} = |z_{\mathrm{M}_{1}}| - |z_{\mathrm{S}_{1}}|$, and $\Delta_{2} := z_{\mathrm{M}_{2}} - z_{\mathrm{S}_{2}} = |z_{\mathrm{M}_{2}}| - |z_{\mathrm{S}_{2}}|$ -- see Fig.~\ref{fig:layout}. The resultant magnetic field due to both current layers, $\mathbold{H} = \mathbold{H}_{1} + \mathbold{H}_{2}$, is considered to be sampled continuously within $\mathrm{M}_{1,2}$.

\subsection{Finding a unique solution for each current layer}
Equation \eqref{eq:B-S} is assumed to apply to each current layer, $\mathrm{S}_{1,2}$. In order to express \eqref{eq:B-S} in such a way that it can be easily inverted to find $\mathbold{J}_{\mathrm{S}_{1,2}}$, it is necessary to neglect the $z$-component of $\mathbold{J}_{\mathrm{S}_{1,2}}$, and to assume that its $x$- and $y$-components do not vary in the $z$-direction. As a preliminary, we consider two $z'$ variables, $z'_{1,2}$, with which to integrate over their respective current layers, and define two additional positive differences: $\Delta_{1}' = z_{1}' - z_{\mathrm{M}_{1}} = |z_{\mathrm{M}_{1}}| - |z_{1}'|$ and $\Delta_{2}' =z_{\mathrm{M}_{2}} - z_{2}' = |z_{\mathrm{M}_{2}}| - |z_{2}'|$. Taking $i = 1,2$, and focusing on the $x$-component of the magnetic field, $H_{x,\mathrm{M}_{i}}$, produced by its nearest current layer, $\mathrm{S}_{i}$, \eqref{eq:B-S} then becomes

\begin{equation}\label{eq:Bx}
    (-1)^{i}H_{x,\mathrm{\mathrm{M}}_{i}}(x,y,\Delta_{i}) = \frac{1}{4\pi}\int_{z_{\mathrm{S}_{i}}-\delta_{i}/2}^{z_{\mathrm{S}_{i}}+\delta_{i}/2}\int_{-\infty}^{+\infty}\int_{-\infty}^{+\infty}\frac{J_{y,\mathrm{S}_{i}}(x',y')\Delta_{i}'}{((x-x')^{2}+(y-y')^{2}+\Delta_{i}'^2)^{3/2}}dx'dy'dz_{i}'.
\end{equation}

\noindent The minus-sign on the left-hand side is introduced by considering the relative positions of $\mathrm{M}_{1,2}$ with respect to $\mathrm{S}_{1,2}$, whilst keeping $\Delta_{i}' > 0$ for clarity.

\subsection{Performing spatial filtering}
By changing the order of integration, \eqref{eq:Bx} may be seen as a convolution in $x$ and $y$ between $J_{y,\mathrm{S}_{i}}$ and the Green function, $G_{i}$, where

\begin{equation}\label{eq:Green}
    G_{i}(x,y,\Delta_{i}) = \frac{1}{4\pi}\int_{z_{\mathrm{S}_{i}}-\delta_{i}/2}^{z_{\mathrm{S}_{i}}+\delta_{i}/2}\frac{\Delta_{i}'}{(x^{2}+y^{2}+\Delta_{i}'^{2})^{3/2}}dz_{i}'.
\end{equation}

\noindent In order to find the current components $J_{y,\mathrm{S}_{i}}$ by inverting \eqref{eq:Bx}, the convolution theorem (in two dimensions) may be employed to convert the integral into a product of two Fourier-transformed functions \cite{Hecht2017}. This two-dimensional Fourier transform ($\mathcal{F}$) pair is defined here as

\begin{subequations}\label{eq:ftPair}
    \begin{align}
        \hat{f}(k_{x},k_{y}) = \mathcal{F}(f(x,y)) = \int\limits_{-\infty}^{+\infty}\int\limits_{-\infty}^{+\infty}f(x,y)\exp(-i(k_{x}x+k_{y}y))dxdy, \\
        f(x,y) = \mathcal{F}^{-1}(\mathcal{F}(f(x,y))) = \frac{1}{(2\pi)^{2}}\int\limits_{-\infty}^{+\infty}\int\limits_{-\infty}^{+\infty}\hat{f}(k_{x},k_{y})\exp(i(k_{x}x+k_{y}y))dk_{x}dk_{y},
        \end{align}
\end{subequations}

\noindent where $k_{x,y}$ are angular wavenumbers, and $\hat{f}(k_{x},k_{y})$ and $f(x,y)$ are bivariate functions resulting from the forward and inverse transforms, respectively \cite{Roth1989}. An exact solution to the two-dimensional Fourier transform of the convolution kernel in \eqref{eq:Green} has been demonstrated by reducing the problem to a Hankel transform \cite{Blakely1995,Jooss1998,Zuber2018,Bracewell2000,Gradstejn2015}. The resulting Fourier space equivalent of \eqref{eq:Bx} is

\begin{subequations}\label{eq:conv}
    \begin{align}
        (-1)^{i}\hat{H}_{x,\mathrm{S}_{i}}(k_{x},k_{y},\Delta_{i}) = \hat{G}(k_{x},k_{y},\Delta_{i})\hat{J}_{y,\mathrm{S}_{i}}(k_{x},k_{y}), \label{subeq:conv} \\
        \hat{G}(k_{x},k_{y},\Delta_{i}) = \frac{1}{2}\int_{z_{\mathrm{S_{i}}}-\delta_{i}/2}^{z_{\mathrm{S_{i}}}+\delta_{i}/2}\exp\left(-\Delta_{i}'\sqrt{k_{x}^{2}+k_{y}^{2}}\right)dz_{i}'. \label{subeq:exp}
    \end{align}
\end{subequations}

\noindent The spatially decaying behaviour here is characteristic of upward continuation -- i.e., moving away from the source of a measured signal \cite{Ravat}. Evaluating the integral over $z_{i}'$, and using $k=\sqrt{k_{x}^{2}+k_{y}^{2}}$, yields

\begin{equation}\label{eq:product}
    (-1)^{i}\hat{H}_{x,\mathrm{M}_{i}}(k_{x},k_{y},\Delta_{i}) = \frac{1}{k}\sinh\left(\frac{k\delta_{i}}{2}\right)\exp(-k\Delta_{i})\hat{J}_{y,\mathrm{S}_{i}}(k_{x},k_{y}).
\end{equation}

\noindent To reiterate, \eqref{eq:product} expresses the magnetic field generated by its respective current layer. It is worth noting that, for $k\delta_{i}/2 << 1$, the hyperbolic sine prefactor reduces to $\delta_{i}/2$, consistent with Refs.~\cite{Lima2006,Roth1989,Bason2022}. 

\noindent By expressing \eqref{eq:Bx} as \eqref{eq:product}, it becomes straightforward to calculate the current density in each current layer in terms of the magnetic field that it produces. It should be noted that a prerequisite for this upward continuation in $z$ to be valid is that there be no current sources within the region through which the continuation is carried out \cite{Blakely1995}. At first glance, this appears to rule out the applicability of this technique to the two-layer problem. However, in this case, upward continuation will only be performed from each current layer independently -- i.e., without accounting for the other layer; it will be shown next that, by considering the resultant fields from both current layers at the measurement planes, a second type of inverse problem may be formed and solved to reconstruct the current densities in each layer. The applicability of this approach will then be corroborated with a numerical experiment. 

\newpage

\subsection{Forming a `continuation matrix'}
Having obtained expressions describing the $\hat{H}_{x,\mathrm{M}_{i}}$ component independently produced by each current layer, $\mathrm{S}_{i}$, which are each suitable for inversion, the next step is to consider the resultant component due to both layers, as detected at each measurement plane. Since magnetic fields sum vectorially, and by the linearity of the Fourier transform, the total Fourier-transformed field component at a measurement plane is simply given by the sum of the Fourier-transformed fields due to each current layer. In order to consider fields detected at both measurement planes, the previous notation needs to be generalised.

For both $i,j = 1,2$, let $\hat{H}_{x,\mathrm{M}_{i}}$ denote the $x$-components of the Fourier-transformed fields detected at measurement planes $\mathrm{M}_{i}$, and let $\hat{J}_{y,\mathrm{S}_{j}}$ denote the $y$-components of the Fourier-transformed current densities of current layers $\mathrm{S}_{j}$. Moreover, let $\hat{G}_{ij}$ represent the Green function associating $\hat{J}_{y,\mathrm{S}_{j}}$ to $\hat{H}_{x,\mathrm{M}_{i}}$. In this way, the problem can be expressed in terms of a `continuation matrix', $\mathbold{\hat{G}}$, as

\begin{equation}\label{eq:jMatrix}
    \mathbold{\hat{H}}_{x} = \mathbold{\hat{G}}\mathbold{\hat{J}}_{y},
\end{equation}

\noindent where $\mathbold{\hat{J}}_{y} = (\hat{J}_{y,\mathrm{S}_{1}},\hat{J}_{y,\mathrm{S}_{2}})^{\intercal}$ and $\mathbold{\hat{H}}_{x} = (-\hat{H}_{x,\mathrm{M}_{1}},\hat{H}_{x,\mathrm{M}_{2}})^{\intercal}$. $\mathbold{\hat{G}}$ is a $2\times 2$ matrix containing all the information about the attenuation of each field away from each current layer:

\begin{equation}
    \mathbold{\hat{G}} =
    \begin{pmatrix}
        \hat{G}_{11} & \hat{G}_{12} \\
        \hat{G}_{21} & \hat{G}_{22}
    \end{pmatrix}.
\end{equation}

\noindent $\hat{\mathbold{G}}$ may then be inverted to solve \eqref{eq:jMatrix} for $\mathbold{\hat{J}}_{y}$. Here, $\mathbold{\hat{G}}$ takes the following form:

\begin{equation}
    \mathbold{\hat{G}} = \frac{1}{k} 
    \begin{pmatrix}
        \sinh(k\delta_{1}/2)\exp(-k\Delta_{1}) & \sinh(k\delta_{2}/2)\exp(-k(\Delta_{1}+\Delta_{\mathrm{S}})) \\
        \sinh(k\delta_{1}/2)\exp(-k(\Delta_{2}+\Delta_{\mathrm{S}})) & \sinh(k\delta_{2}/2)\exp(-k\Delta_{2})
    \end{pmatrix}.
\end{equation}

\noindent $\Delta_{1,2}$ each denote the distance between a measurement plane and its nearest current layer, and $\Delta_{\mathrm{S}}$ is the spacing between the two current layers. Note that upward continuation has been assumed in the off-diagonal elements in assuming further attenuation of the field over $\Delta_{\mathrm{S}}$. It is worth emphasising here again that each current layer is treated independently -- i.e., the upward continuation involved is not compromised by the presence of the other current layer.

\subsection{Reconstructing the current densities in each layer}
Solution of \eqref{eq:jMatrix} for the $\mathbold{\hat{J}}_{y}$ components in each current layer gives the following:

\begin{subequations}\label{eq:j_y}
    \begin{align}
        \hat{J}_{y,\mathrm{S}_{1}} = -\frac{k}{2}\csch\left(\frac{k\delta_{1}}{2}\right)\csch(k\Delta_{\mathrm{S}})(\exp(k(\Delta_{1} + \Delta_{\mathrm{S}}))\hat{H}_{x,\mathrm{M}_{1}} + \exp(k\Delta_{2})\hat{H}_{x,\mathrm{M}_{2}}), \label{subeq:j_yS1} \\
        \hat{J}_{y,\mathrm{S}_{2}} = \frac{k}{2}\csch\left(\frac{k\delta_{2}}{2}\right)\csch(k\Delta_{\mathrm{S}})(\exp(k(\Delta_{2} + \Delta_{\mathrm{S}}))\hat{H}_{x,\mathrm{M}_{2}} + \exp(k\Delta_{1})\hat{H}_{x,\mathrm{M}_{1}}). \label{subeq:j_yS2}
    \end{align}
\end{subequations}

\noindent The corresponding expressions for $\hat{J}_{x}$ may be found by making the replacements $\hat{J}_{y}\rightarrow \hat{J}_{x}$ and $\hat{H}_{x}\rightarrow -\hat{H}_{y}$. Taking the inverse Fourier transform yields the corresponding component of the current density in terms of $x$ and $y$. The spatially growing terms are characteristic of downward continuation -- moving towards the source of a measured signal \cite{Ravat}. In this way, by taking measurements in one plane above and one plane below the current stack, the current densities in each current layer can be reconstructed exactly.

\begin{figure}
    \centering
    \includegraphics[width=\textwidth]{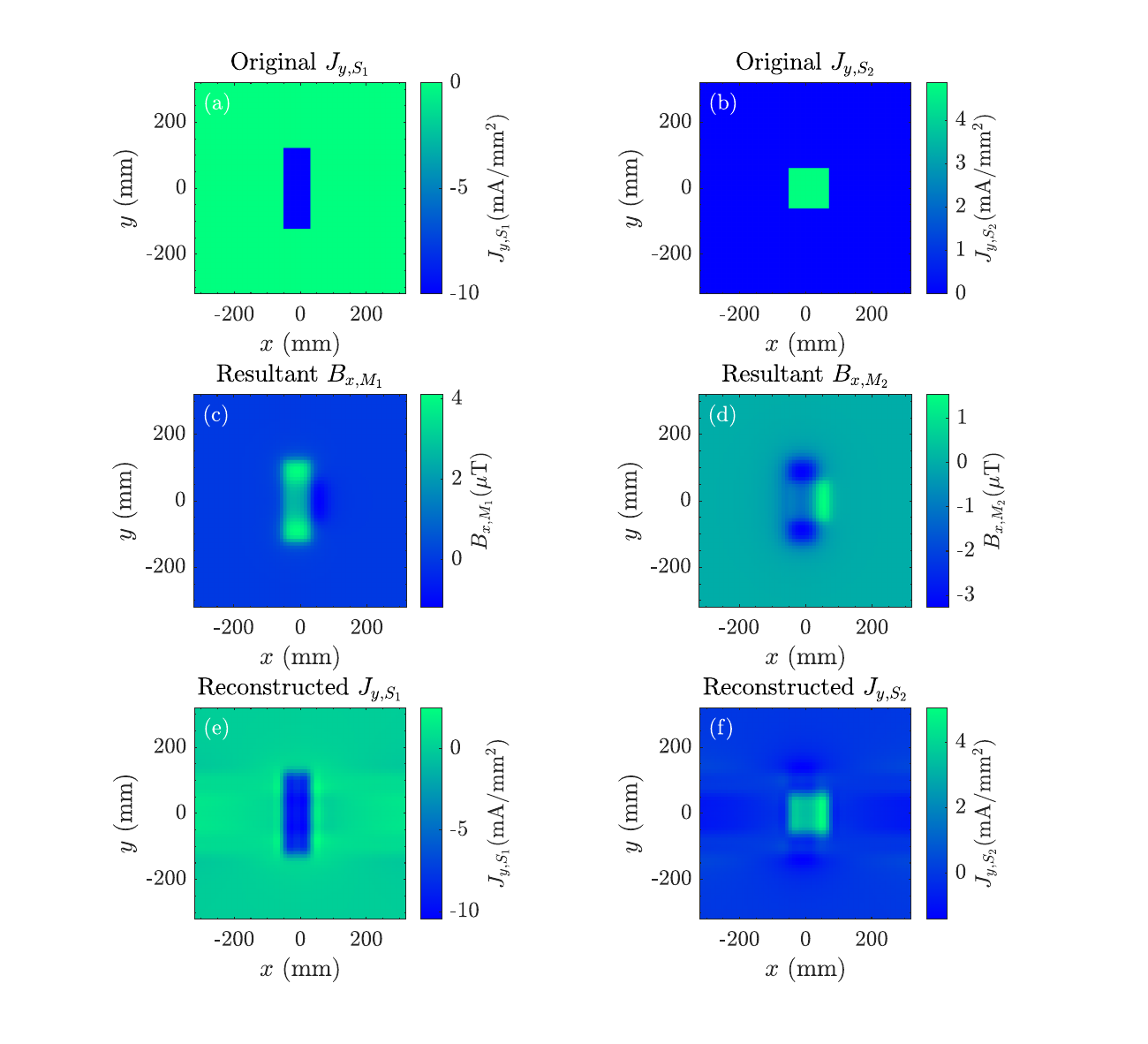}
    \caption{Results from a numerical simulation, in MATLAB, of the reconstruction of current densities in two layers, from magnetic flux densities in two different measurement planes -- one above the current stack and one below. (a) and (b) show the original $y$-components of the current densities, $J_{y, \mathrm{S_{1,2}}}$, in layers $\mathrm{S}_{1,2}$. (c) and (d) show the resultant $x$-components of the magnetic flux densities, $B_{x, \mathrm{M_{1,2}}}$, in planes $\mathrm{M}_{1,2}$; the contributions from the two layers partially overlap. (e) and (f) show the reconstructed $J_{y, \mathrm{S_{1,2}}}$, calculated by taking fast Fourier transforms (FFTs) of $B_{x, \mathrm{M_{1,2}}}$, then using \eqref{eq:j_y}, before taking inverse FFTs of the result. Blackman-Harris and Hann functions were used to mitigate windowing effects caused by the FFTs used to implement this method, although some small aberrations remain, which increase the range of values for the reconstructed current densities. The currents (all in the $y$-direction) through $\mathrm{S}_{1}$ and $\mathrm{S}_{2}$ were $-0.8$ A and 0.7 A, respectively; the respective widths of the layers (in the $x$-direction) were 80 mm and 120 mm. The other parameters were: $\Delta_{1} = 14$ mm, $\Delta_{2} = 13$ mm, $\Delta_{\mathrm{S}} = 11$ mm, $\delta_{1} = 1$ mm, and $\delta_{2} = 1.2$ mm.}
    \label{fig:twoSheets}
\end{figure}

As a numerical demonstration of \eqref{eq:j_y}, two vertically separated current layers -- and components of their resultant magnetic flux densities, $B_{x;\mathrm{M}_{1,2}}$, at two different measurement planes, $\mathrm{M}_{1,2}$ -- were simulated in MATLAB. Magnetic flux density, as the experimentally accessible quantity, was simulated, rather than the magnetic field; it was assumed that $B_{x;\mathrm{M}_{1,2}} = \mu_{0}H_{x;\mathrm{M}_{1,2}}$, where $\mu_{0}$ is the vacuum permeability. The current density in each layer was then reconstructed from the magnetic flux densities using \eqref{eq:j_y}. The results are illustrated in Fig.~\ref{fig:twoSheets}.

\newpage

\subsection{Non-invasive measurements in terms of the continuation matrix}
Non-invasive measurements here correspond to measurement planes being placed outside of the current stack. A necessary but insufficient condition for $\hat{\mathbold{G}}$ to be invertible is that the number of measurement planes be equal to the number of current layers, so that $\hat{\mathbold{G}}$ is a square matrix. At each measurement plane, knowledge of the field components due to each current layer is lost, since only their sum is known. This can be compensated in the case of two current layers by placing one measurement plane above the current stack and one below, since the resulting system of equations is linearly independent. If, however, both measurement planes are placed on the same side of the current stack, then the resulting system of equations is linearly dependent, so the associated continuation matrix is rank-deficient, making the problem non-invertible. 

More generally, if more than two current layers are considered, then the requirement of matching the number of measurement planes to the number of current layers implies more than one measurement plane on one or both sides of the current stack, resulting in a non-invertible problem. This suggests two approaches -- one invasive and one non-invasive: The invasive method is to intercalate measurement planes between current layers such that the problem is invertible. The non-invasive option involves adopting a more general definition of matrix inversion \cite{Penrose1955,Barata2011}, which would allow for non-matched numbers of current layers and measurement planes, but at the expense of the accuracy of the approximate reconstruction. It would then be a matter of refining this reconstruction using realistic prior assumptions about the layered structure of the current-carrying stack in question. The latter approach, based on generalised inversion of the associated continuation matrix, is to be considered in future work.

\section{Summary}
We have demonstrated calculations that reconstruct electric current densities in two stacked layers, based on knowledge of their resultant magnetic fields in two planes -- one above the stack, one below. After deriving the relevant system of equations, the efficacy of the reconstruction was demonstrated in a numerical experiment. This showed that, aside from small aberrations due to windowing effects arising from FFTs, the current densities in the two layers were reconstructed accurately. In working with the magnetic flux densities in this numerical demonstration, the vacuum permeability was used -- i.e., it was assumed that no significantly magnetic materials were present. This would be consistent with aluminium or copper current collectors in certain types of battery cell, for example \cite{Garayt2023,Murray2019,Pathan2019}. As well as the non-invasive characterisation of battery cells, this work is also anticipated to be useful for imaging electrical activity in printed circuit boards, capacitors, and fuel cells. In addition to applying generalised inversion and regularisation schemes to the continuation matrices associated with multi-layer magnetic inverse problems, future work will also consider the consequences of discretisation of the measurement plane(s), in the form of one or more regular arrays of magnetometers.

\section{Acknowledgements}
This work was supported by the UK Quantum Technologies Hub for Sensors and Timing (EPSRC Grant EP/T001046/1), University of Sussex Strategic Development Fund, and Innovate UK: Batteries—ISCF 42186 Quantum sensors for end-of-line battery testing.

\section{Data availability statement}
No new data were created or analysed in this study.

\newpage

\bibliographystyle{ieeetr}
\bibliography{two-layer}

\end{document}